\documentclass[12pt,a4 paper]{article}
\usepackage[dvips]{graphicx}
\begin{document}
\title{ON FRACTAL DIMENSION OF PROTON AT SMALL $x$}
\author{D.K.Choudhury and Rupjyoti Gogoi\\Gauhati University}
\maketitle
\begin{abstract}
Recently the concept of self similarity in the structure of the proton at small x has been introduced.We estimate the fractal dimension of proton in analogy with classical monofractals.\\

\vspace{4cm}
\end{abstract}
\newpage
\section{Introduction}
	
   Self similarity is a familiar property in nature[1,2]. Many of the seemingly irregular shapes of nature have hidden self similarity in them.It is not the usual symmetry with respect to rotation or translation,but symmetry with respect to scale or size: a small part of a system is self similar to the entire system. Such a system is defined through its self similar dimension, which is in general fractal, hence called fractal dimension. Classical fractals discussed in standard references[1,2] are Cantor Dust, Koch Curve and Sierpinski gasket whose fractional dimensions are 0.63,1.26 and 1.585 respectively, which lie between Euclidean point and surfaces.\\

Notion of self similarity and fractal dimensions are being used in the phase spaces of hadron multiparticle production processes since nineteen eighties[3-7].However these ideas did not attract much attention in contemporary physics of deep inelastic lepton hadron scattering, due to its obvious lack of applicability. Although a related approach employing self organized criticality [8], has been pursued in reference [9-12] for quark-gluon systems . Only recently [13], Lastovicka has developed relevant formalism and applied it to deep inelastic electron proton scattering at low $x$ and proposed a functional form of the structure function $F_{2}(x,Q^{2})$. Specifically a description of $F_{2}(x,Q^{2})$ reflecting self similarity is proposed with a few parameters which are fitted to recent HERA data[14,15].The specific parameterization is claimed to provide an excellent description of the data which covers a region of four momentum transferred squared $0.45<Q^{2}\le120GeV^{2}$ and of Bjorken $x$,$6.2\times10^{-7}<x\le0.01$.\\

More recently, it was observed [16] that the positivity of fractal dimensions prohibits some of the fitted parameters of the structure function of reference[13]. Specifically out of the fractal dimensions $D_{1}$,$D_{2}$ and $D_{3}$ one ($D_{3}$)was estimated to be negative ($D_{3}\sim-1.3$). However the positivity of fractal dimension forbids such negative value.\\

In order to avoid such possibility, we suggest that the proton is described by the single self similarity dimension D. This then facilitates one to compare the self similarity nature of the proton at low $x$ with the classical multifractals, which is the aim of the present paper. In section2, we report the essentials of the formalism while section 3 contains the conclusion.
\section{Formalism}
Under the hypothesis of self similarity of the proton structure at low $x$, Lastovicka[13] obtained the following form of the structure function $F_{2}(x,Q^{2})$, 
\begin{equation}
F_{2}(x,Q^{2})=\frac{e^{D_{0}}(\frac{Q_{0}^{2}}{\Lambda^{2}})x^{-D_{2}+1}}{1+D_{3}+D_{1}\log(\frac{1}{x})}[x^{-D_{1}\log(1+\frac{Q^{2}}{Q_{0}^{2}})}(1+\frac{Q^{2}}{Q_{0}^{2}})^{D_{3}+1}-1]
\end{equation}
where $D_{1}$,$D_{2}$,$D_{3}$ are the self similarity dimensions associated with the magnification factors $\log(\frac{1}{x})$, $\log(1+\frac{Q^{2}}{Q_{0}^{2}}).\log(\frac{1}{x})$ and $\log(1+\frac{Q^{2}}{Q_{0}^{2}})$, while $D_{0}$ defines the normalization constant.Since the magnification factors should be positive, non-zero and dimensionless, a choice $1+\frac{Q^{2}}{Q_{0}^{2}}$, rather than $Q^{2}$ has been made, while $Q_{0}^{2}$ is arbitrary small virtuality,$Q^{2}>Q_{0}^{2}$. An additional parameter $\Lambda$ has been introduced in to make the structure function dimensionless, a point overlooked in [13]. Explicit confrontation with HERA data [14,15] yields, $D_{0}=0.339\pm0.145$,$D_{1}=0.073\pm0.001$,$D_{2}=1.013\pm0.01$,$D_{3}=-1.287\pm0.01$ and $Q_{0}^{2}=0.062\pm0.01GeV^{2}$. 
As the self similarity dimensions of all known fractals are positive[1,2], by its definitions one expects $D_{1}\ge0$,$D_{2}\ge0$,$D_{3}>0$, a feature overlooked in the empirical fit of [13] as far as $D_{3}$ is concerned.Further in analogy with other classic fractals [1,2] we also assume that proton at low $x$ is a monofractal with just one single fractal dimension, so that 
\begin{equation}
D_{1}=D_{2}=D_{3}=D
\end{equation}
Under such a hypothesis, equation (1) is rewritten as,
\begin{equation}
F_{2}(x,Q^{2})=[\frac{e^{D_{0}}(\frac{Q_{0}^{2}}{\Lambda^{2}})x^{-D+1}}{1+D+D\log\frac{1}{x}}][(\frac{1}{x})^{D\log(1+\frac{Q^{2}}{Q_{0}^{2}})}(1+\frac{Q_{2}}{Q_{0}^{2}})^{D+1}-1]
\end{equation}
For very low $x$,
$\log(\frac{1}{x})>>D$,1
\begin{equation}
F_{2}(x,Q^{2})=\frac{e^{D_{0}}(\frac{Q_{0}^{2}}{\Lambda^{2}})x^{-D+1}}{D\log(\frac{1}{x})}(\frac{1}{x})^{D\log(1+\frac{Q^{2}}{{Q_{0}^{2}}})}(1+\frac{Q^{2}}{Q_{0}^{2}})^{D+1}
\end{equation}
In (4), $D_{0}$ is the only flavor dependent factor, varying for different quark combinations. An immediate prediction of (4) is the slope of the structure function $\lambda_{eff}$ defined through
\begin{equation}
F_{2}(x,Q^{2})\sim Cx^{-\lambda_{eff}(Q^{2})}
\end{equation}
yielding
\begin{equation}
\lambda_{eff}=D[1+\log(1+\frac{Q^{2}}{Q_{0}^{2}})]-1+\frac{\log(\frac{1}{\log(1/x)})}{\log(\frac{1}{x})}
\end{equation}
Recently HERA data at low $x$ [17] have been parameterized through (5) with
\begin{equation}
\lambda_{eff}=a\log(\frac{Q^{2}}{\Lambda^{2}})
\end{equation}
where $a=0.048\pm0.0013(stat)\pm0.0037(syst)$,$c\sim0.18$ and $\Lambda=292\pm20(stat)\pm51(syst)MeV$ consistent with QCD expectation [18,19] above Regge regime[20].
Equating (6) and (7), we obtain the fractal dimension of proton to be,
\begin{equation}
D(x,Q^{2})=\frac{1+a\log(\frac{Q^{2}}{\Lambda^{2}})+\frac{\log(\log(\frac{1}{x}))}{\log(\frac{1}{x})}}{1+\log(1+\frac{Q^{2}}{Q_{0}^{2}})}
\end{equation}
Equation (8) is the main result of the present paper.
\section{Results}

In figure 1-4, we plot $D(x,Q^{2})$ versus $x$[equation(8)] for fixed $Q^{2}$ values,$Q^{2}$=5,8.5, 20, 35, 45, 60, 90, 150 $GeV^{2}$. It shows that the fractal dimension $D(x,Q^{2})$ increases with $x$ but decreases for higher values of $x$ ($x\ge3.9\times10^{-3}$). This decreasing nature of $D(x,Q^{2})$ for higher $x$ is more prominent for high $Q^{2}$. It has also been observed that at high $Q^{2}$,$D(x,Q^{2})$ decreases very slowly instead of increasing.\\

In figure 5-8, we plot $D(x,Q^{2})$ [equation(8)] versus $Q^{2}$ for representative small $x$, $x=1.3\times10^{-4}$(0.00013),$2\times10^{-4}$(0.0002),$3.2\times10^{-4}$(0.00032),$5\times10^{-4}$(0.0005),$8\times10^{-4}$(0.0008),$1.3\times10^{-3}$(0.0013),$2\times10^{-3}$(0.002) and $1.3\times10^{-2}$(0.013). It shows invariable decrease of $D(x,Q^{2})$ as $Q^{2}$ increases. In table 1,we also record the values of $D(x,Q^{2})$ for a few representative data points.Taking the average of the total 148 $x$,$Q^{2}$ data points of H1[12]in the range $1.5\le Q^{2}\le 150GeV^{2}$ and $3.2\times10^{-5}\le x\le 0.2$, we  obtain average fractal dimension to be $D\sim 0.64$. Our analysis thus suggests that if proton is viewed as a monofractal at low $x$,HERA data[14-16] constrains its average dimension $D$ such that in its self similarity nature, it is close to Cantor dust ($D\sim0.63$).\\

Let us conclude this work with a few comments.Replacing the three fractal dimensions of Reference[13] by a continuous function of two variables, equation (8) is a considerable extension of the parameter space and needs close scrutiny. A much more modest but meaningful approach would have been perhaps to make the $D_{3}$ of reference [13] vanishingly small $D_{3}\sim 0$ and see how it differs the original analysis. Furthermore notion of fractal dimension of proton needs clearer physical interpretation.Lastly, the formalism developed here needs direct comparison with structure function data. These aspects are currently under study.\\

It is however tempting to speculate that the notion of monofractal versus multifractal craracterizations of systems pursued actively in other branches of sciences [21] might be of relevence even in the study of structure functions at low $x$.
\section{Acknowledgement}
Both of us thank Dr.Subir Sarkar for helpful discussions. One of us (DKC) acknowledges useful coresspondence with Dr.T.Lastovicka.He also acknowledges warm hospitality at the Institute of Mathematical Sciences, Chennai where a part of the work has done.

\begin{tabular}{||c|c|c||}\hline\hline
\hline
\multicolumn{3}{|c|}{TABLE1:Values of $D(x,Q^{2})$ for a few representative data points}
\\\hline
\emph{$Q^{2}$} &x        &{\emph{$D(x,Q^{2})$}}\\\hline
5 	& $8.18\times10^{-5}$	     &0.754 \\\cline{2-3}
	& $1.30\times10^{-2}$        &0.752 \\\hline
8.5     & $1.39\times10^{-2}$        &0.684 \\\cline{2-3}
        & $2.00\times10^{-2}$        &0.675 \\\hline
20      & $2.68\times10^{-4}$        &0.588 \\ \cline{2-3}
        & $3.20\times10^{-2}$        &0.57  \\\hline
35      & $5.74\times10^{-4}$        &0.537  \\\cline{2-3}
        & $8.00\times10^{-2}$        &0.487 \\\hline
45      & $1.30\times10^{-3}$        &0.517 \\\cline{2-3}
        & $8.00\times10^{-2}$        &0.468 \\\hline
60      & $2.00\times10^{-3}$        &0.496 \\\cline{2-3}
        & 0.013                      &0.491 \\\hline
90      & $3.2\times10^{-3}$         &0.468 \\\cline{2-3}
        & 0.13                       &0.39 \\\hline
150     &$2.00\times10^{-2}$         &0.43 \\\cline{2-3}
        &0.2                         &0.31 \\\hline\hline

\end{tabular}
\newpage
\begin{figure}[t]
\centerline{
\includegraphics[width=10cm]{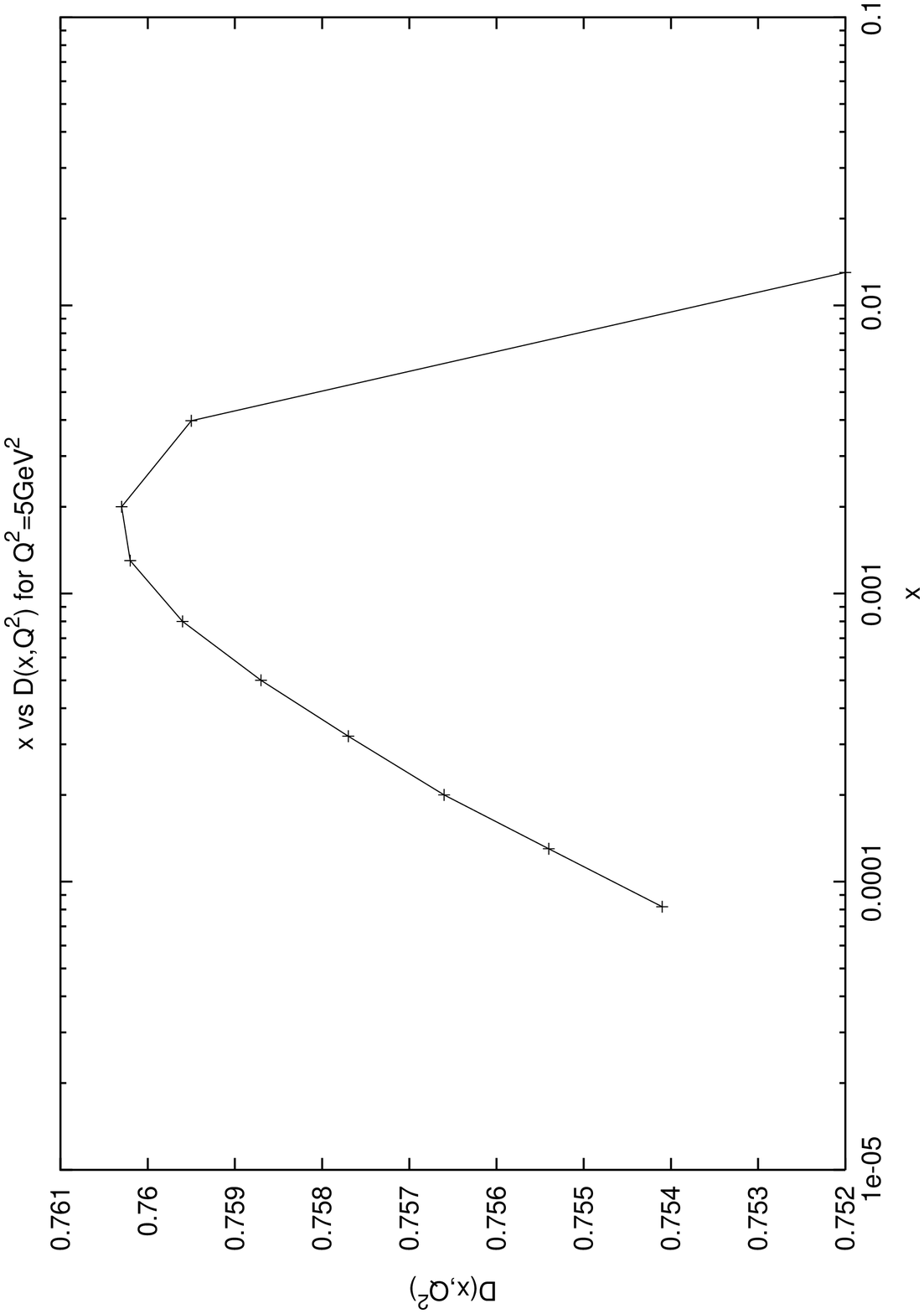}
\includegraphics[width=10cm]{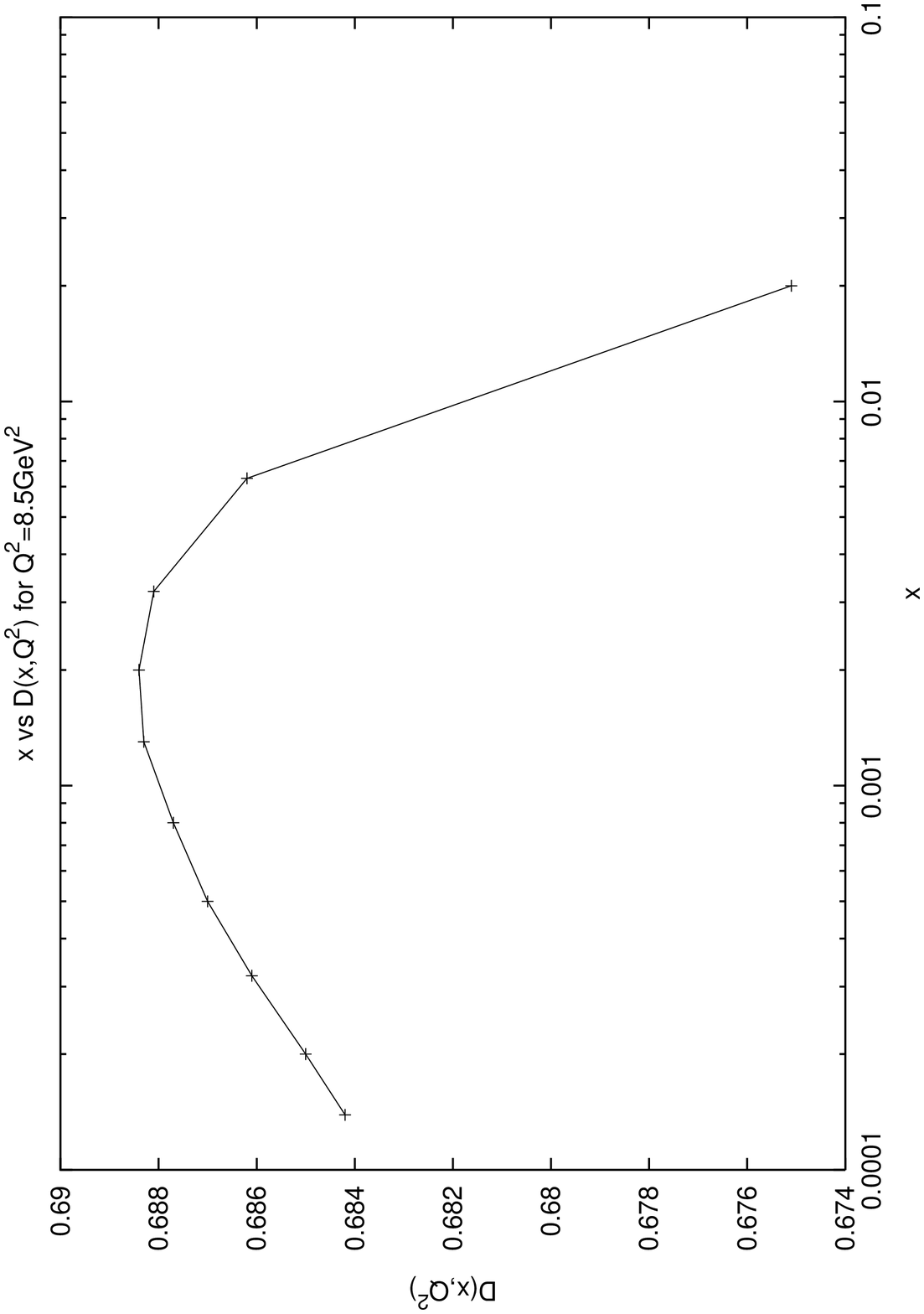}
}
\caption{$x$ versus $D(x,Q^{2})$for $Q^{2}=5GeV^{2}$ and $8.5GeV^{2}$ }
\end{figure}

\begin{figure}[t]
\centerline{
\includegraphics[width=10cm]{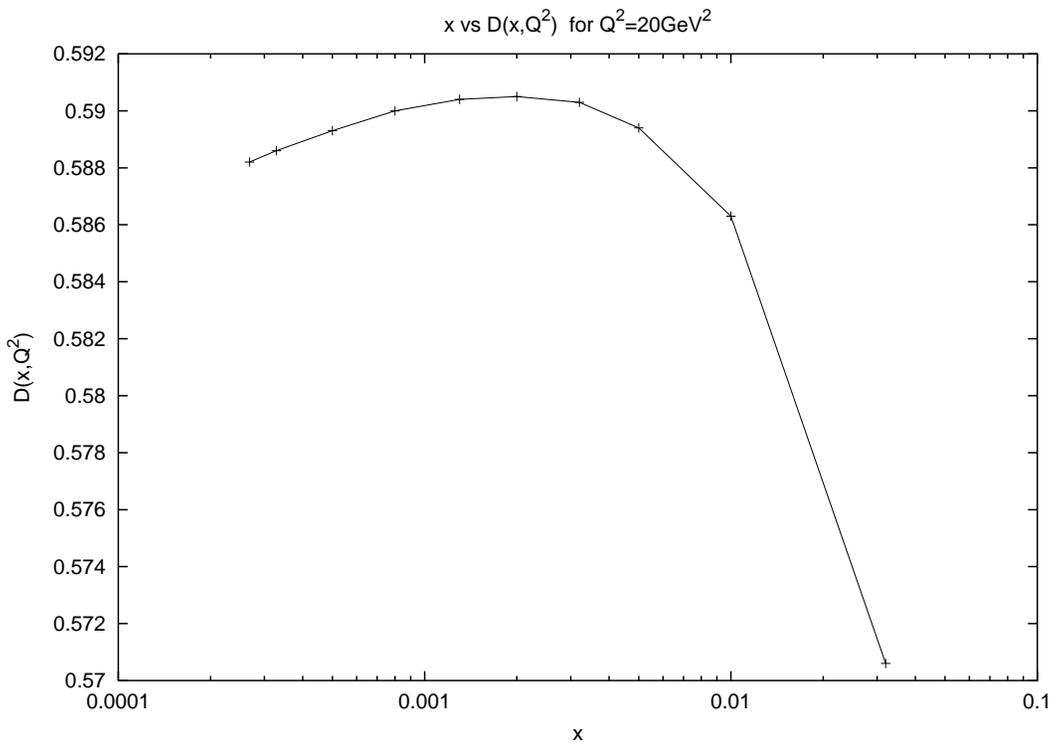}
\includegraphics[width=10cm]{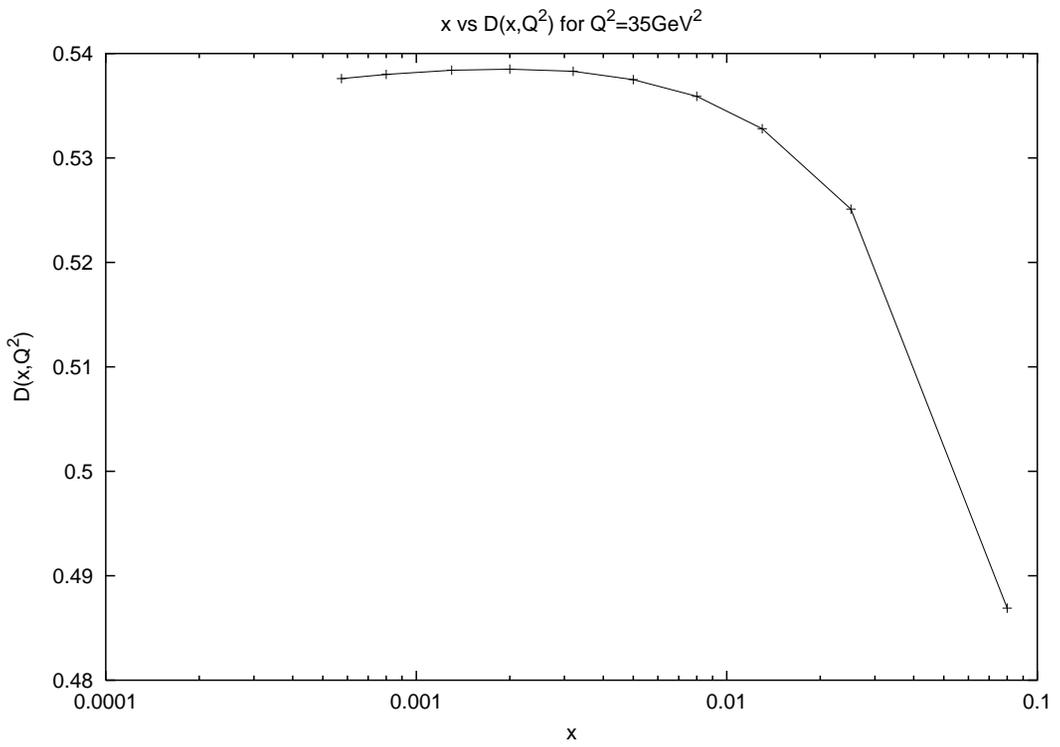}
}
\caption{$x$ versus $D(x,Q^{2})$ for $Q^{2}$=20 and 35 $GeV^{2}$  }
\end{figure}
\begin{figure}[t]
\centerline{
\includegraphics[width=10cm]{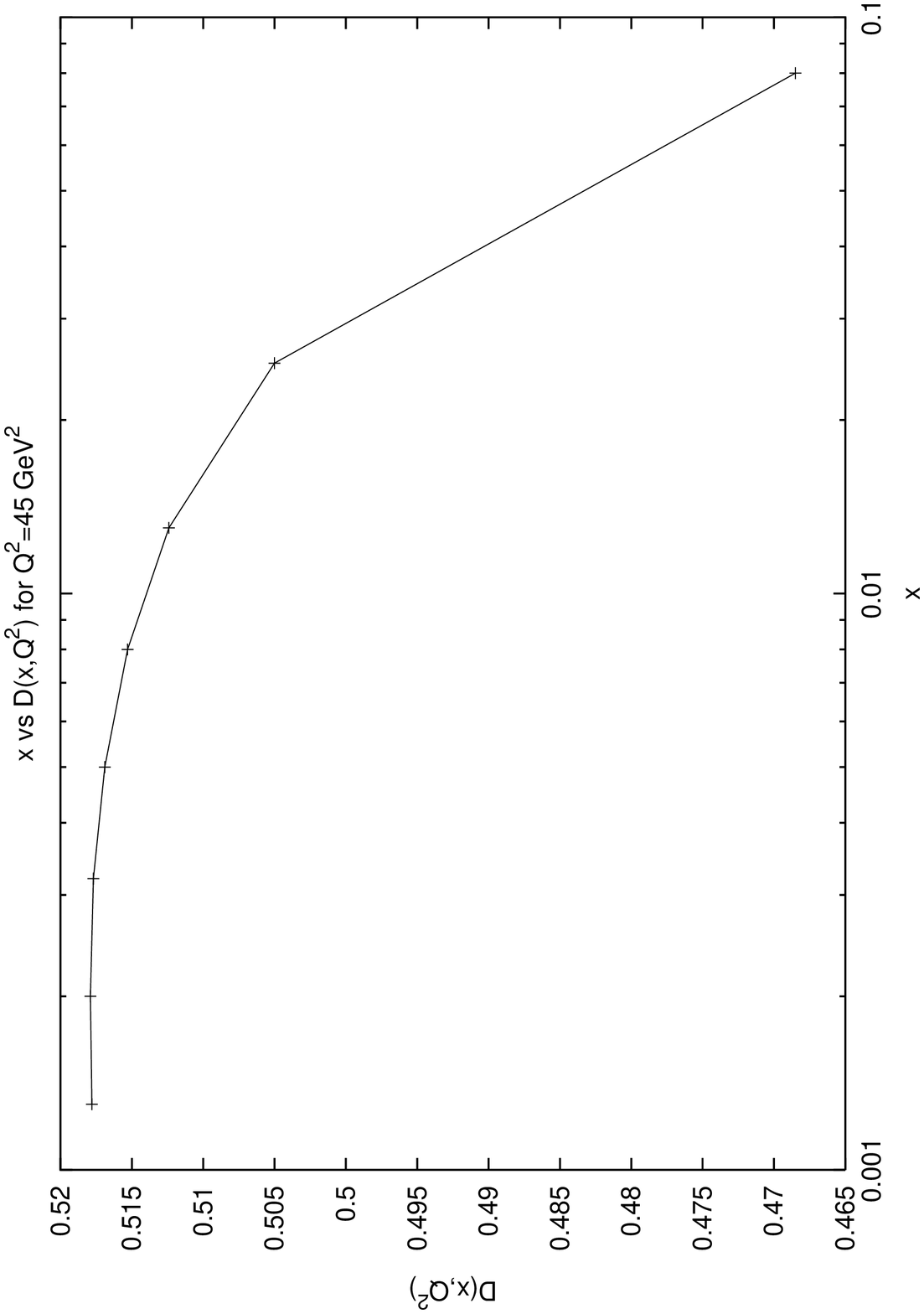}
\includegraphics[width=10cm]{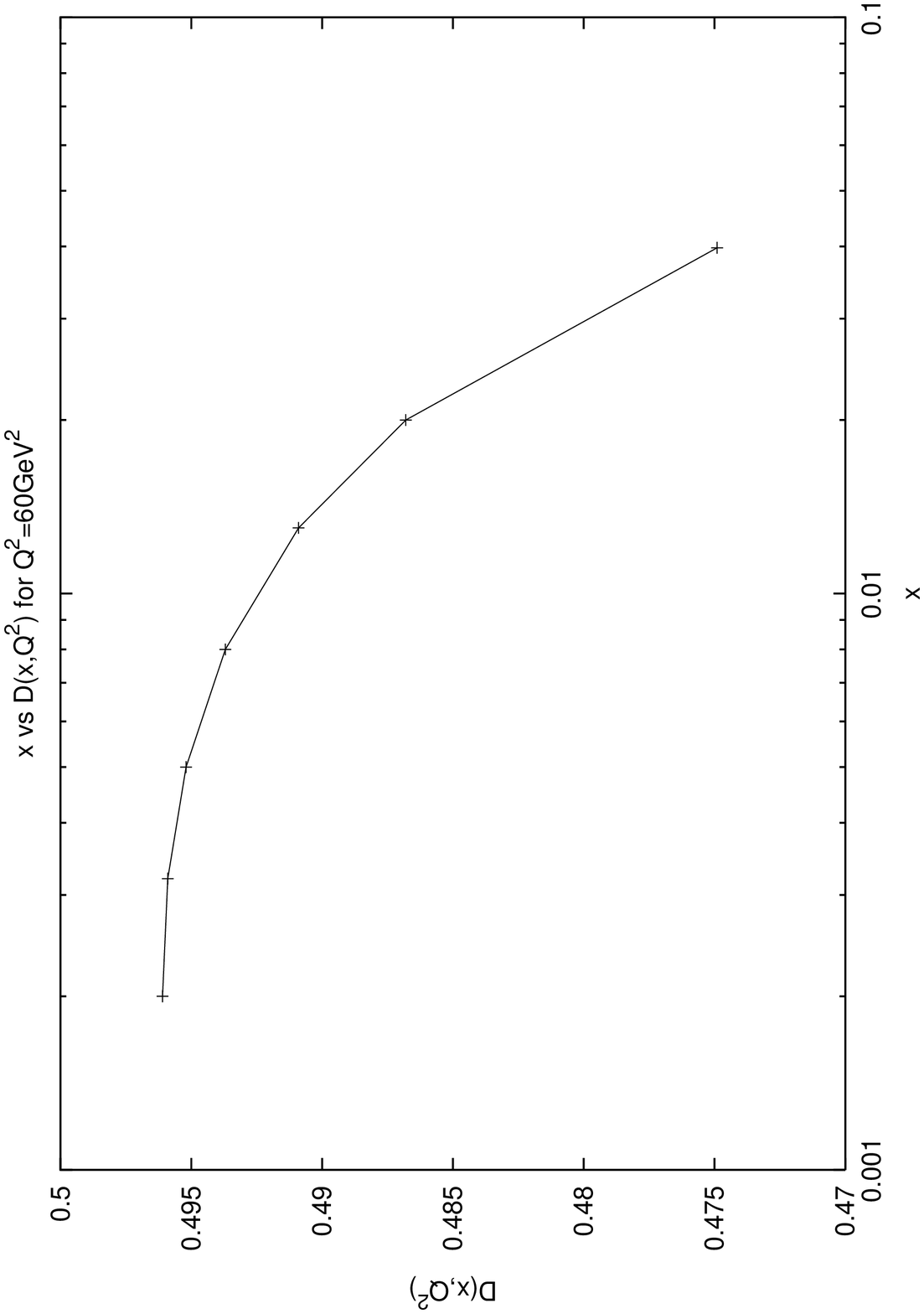}
}
\caption{$x$ versus $D(x,Q^{2})$ for $Q^{2}$=45 and 60 $GeV^{2}$ }
\end{figure}
\begin{figure}[t]
\centerline{
\includegraphics[width=10cm]{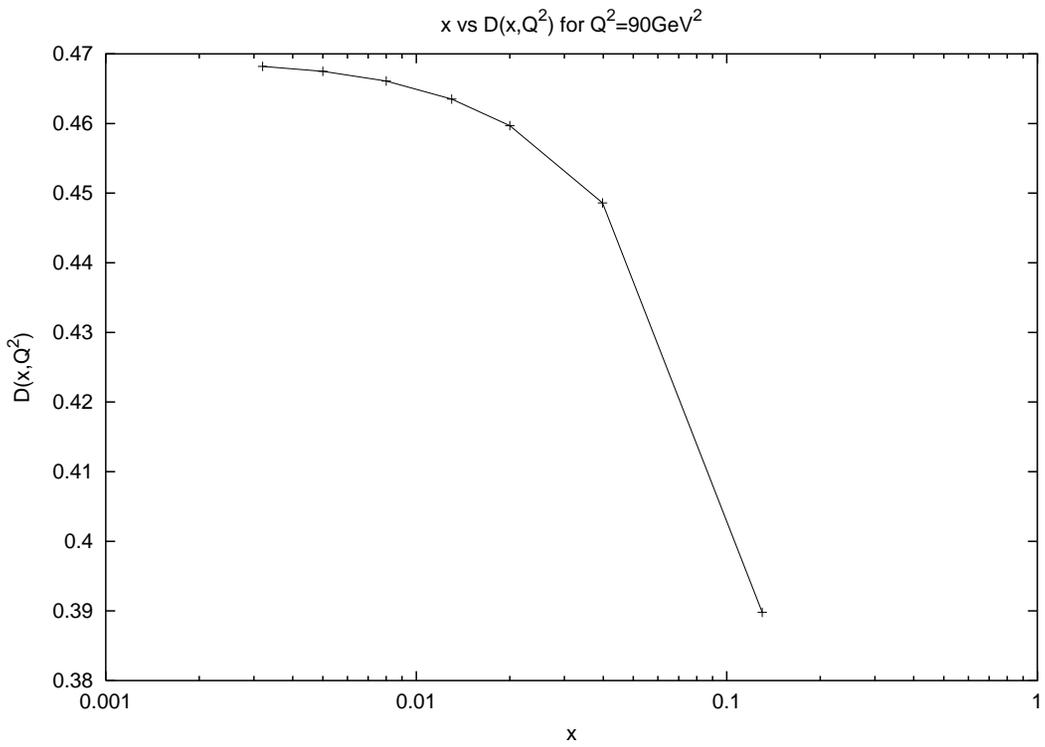}
\includegraphics[width=10cm]{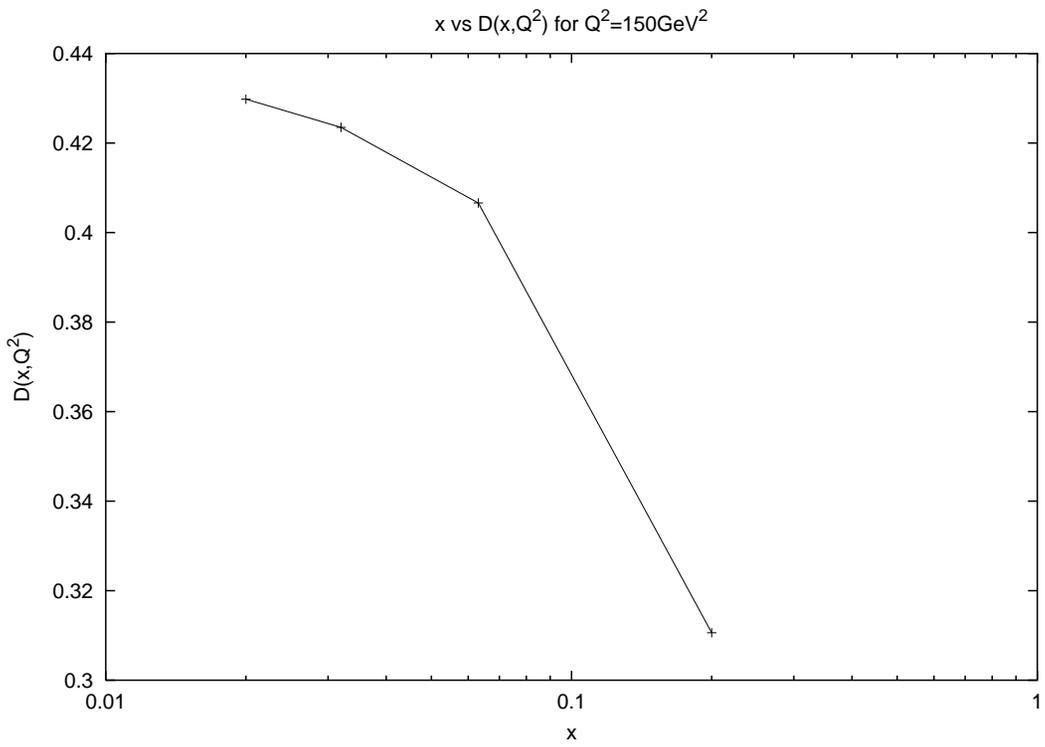}
}
\caption{$x$ versus $D(x,Q^{2})$ for $Q^2$=90 and 150 $GeV^{2}$ }
\end{figure}
\begin{figure}[t]
\centerline{
\includegraphics[width=10cm]{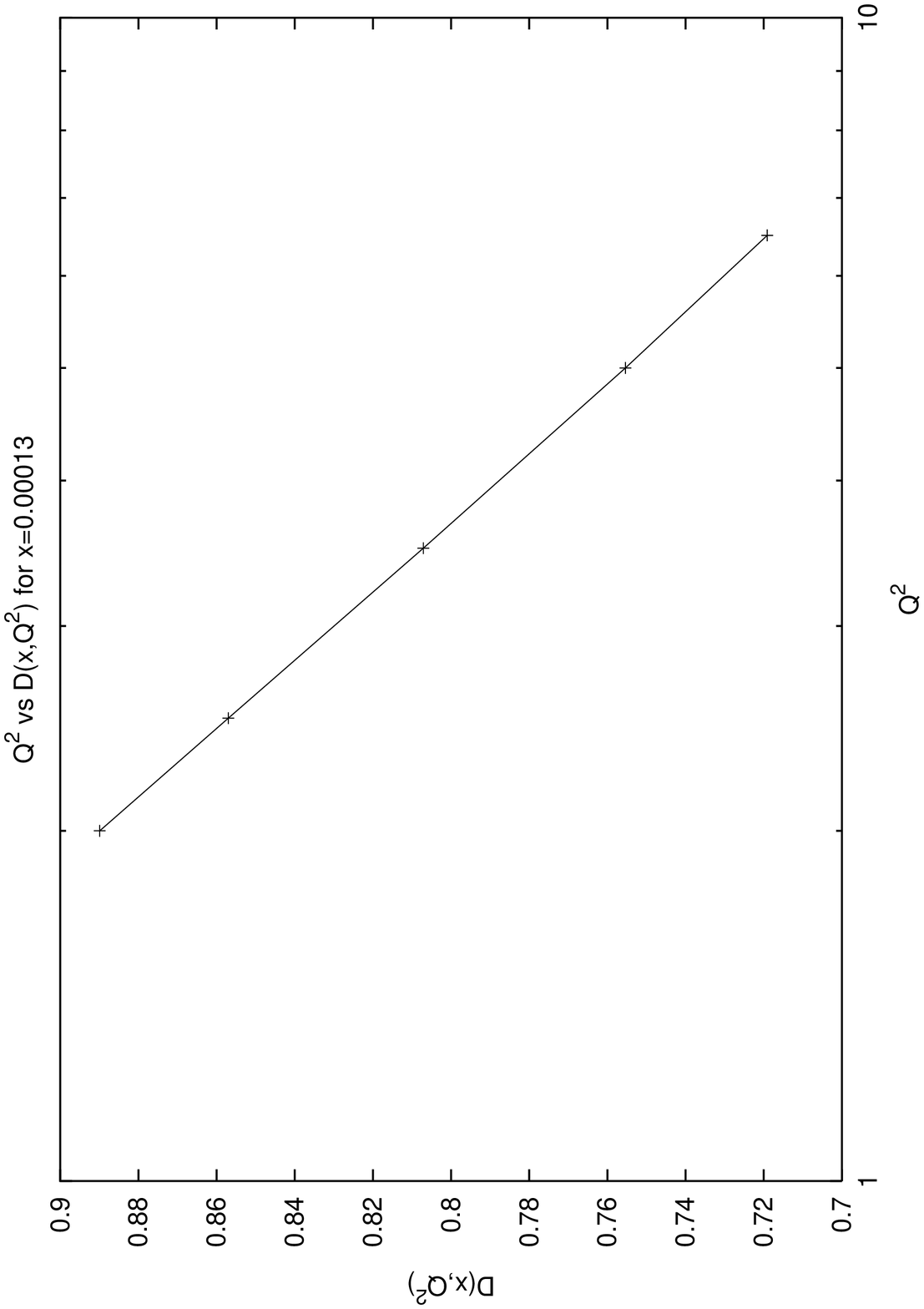}
\includegraphics[width=10cm]{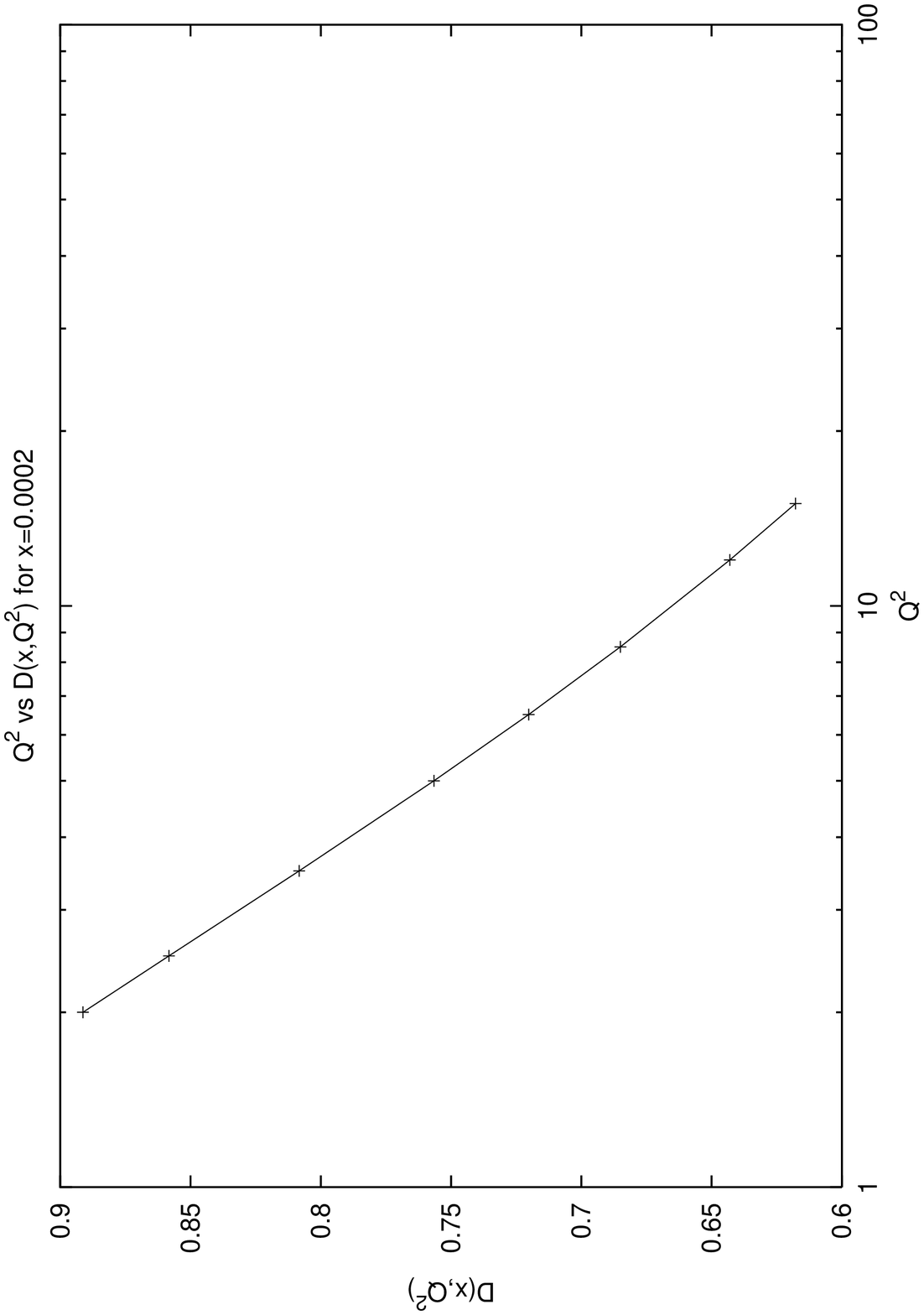}
}
\caption{$Q^{2}$ versus $D(x,Q^{2})$ for $x$=0.00013 and 0.0002}
\end{figure}
\begin{figure}[t]
\centerline{
\includegraphics[width=10cm]{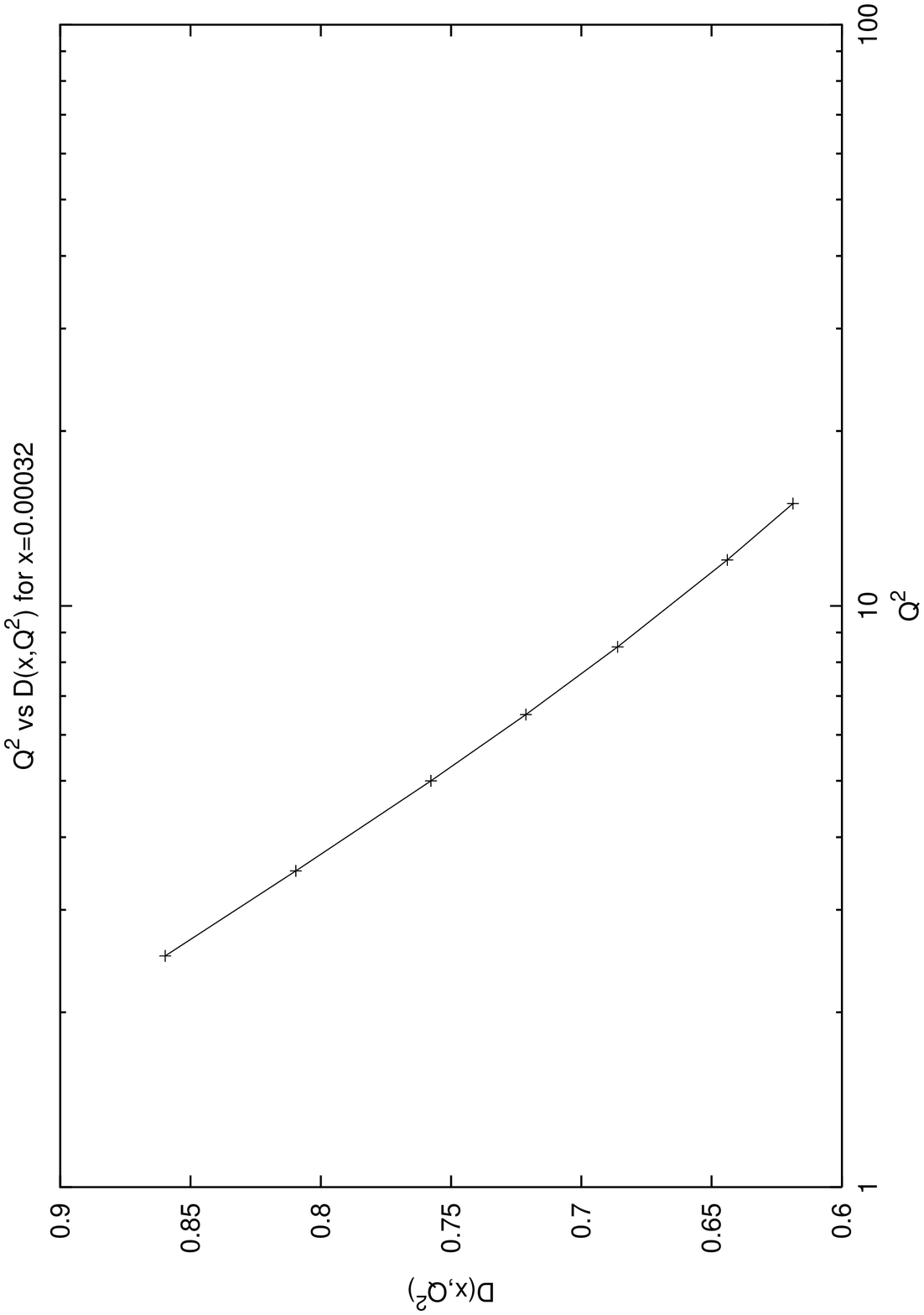}
\includegraphics[width=10cm]{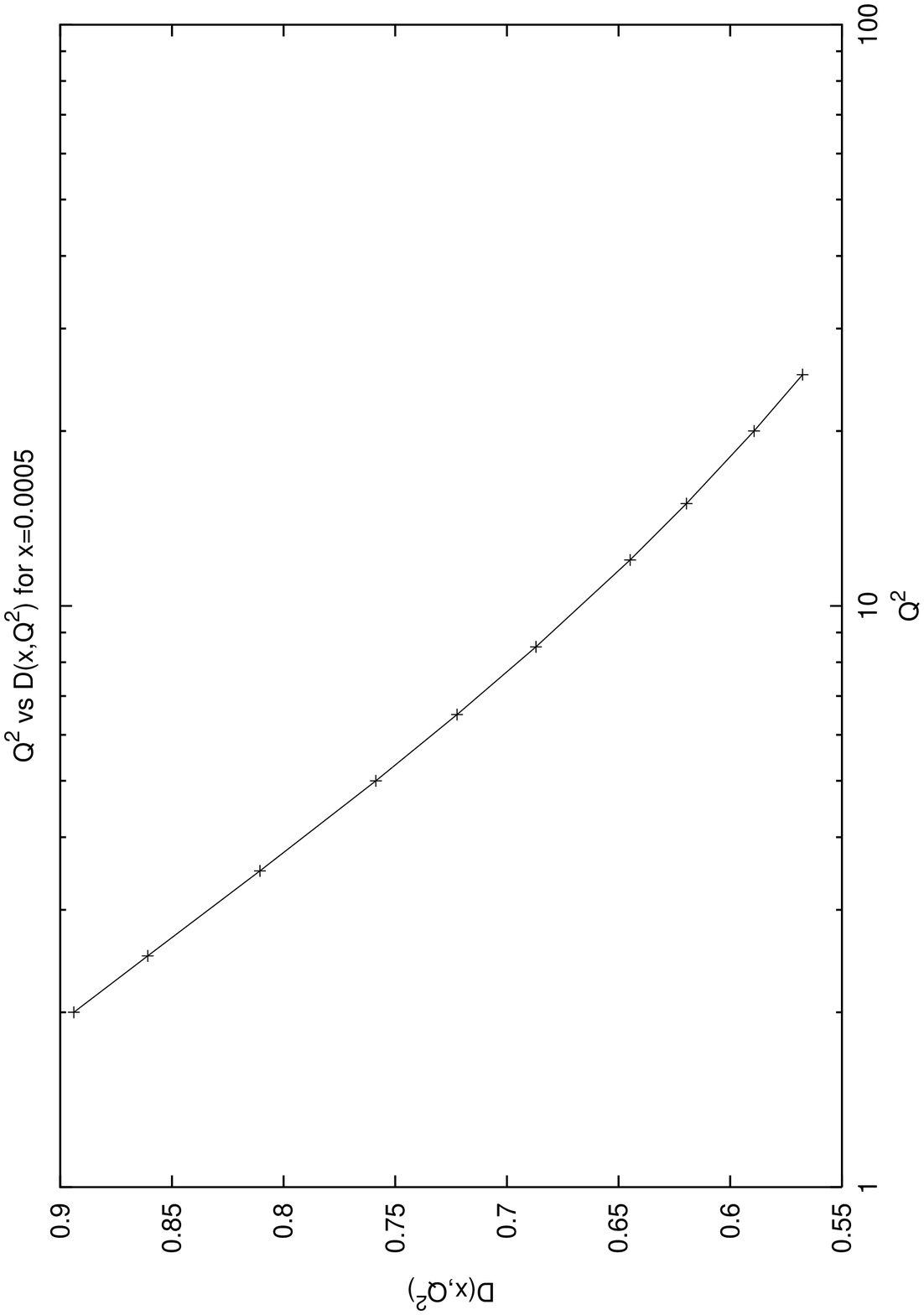}
}
\caption{$Q^{2}$ versus $D(x,Q^{2})$ for $x$=0.00032 and 0.0005}
\end{figure}
\begin{figure}[t]
\centerline{
\includegraphics[width=10cm]{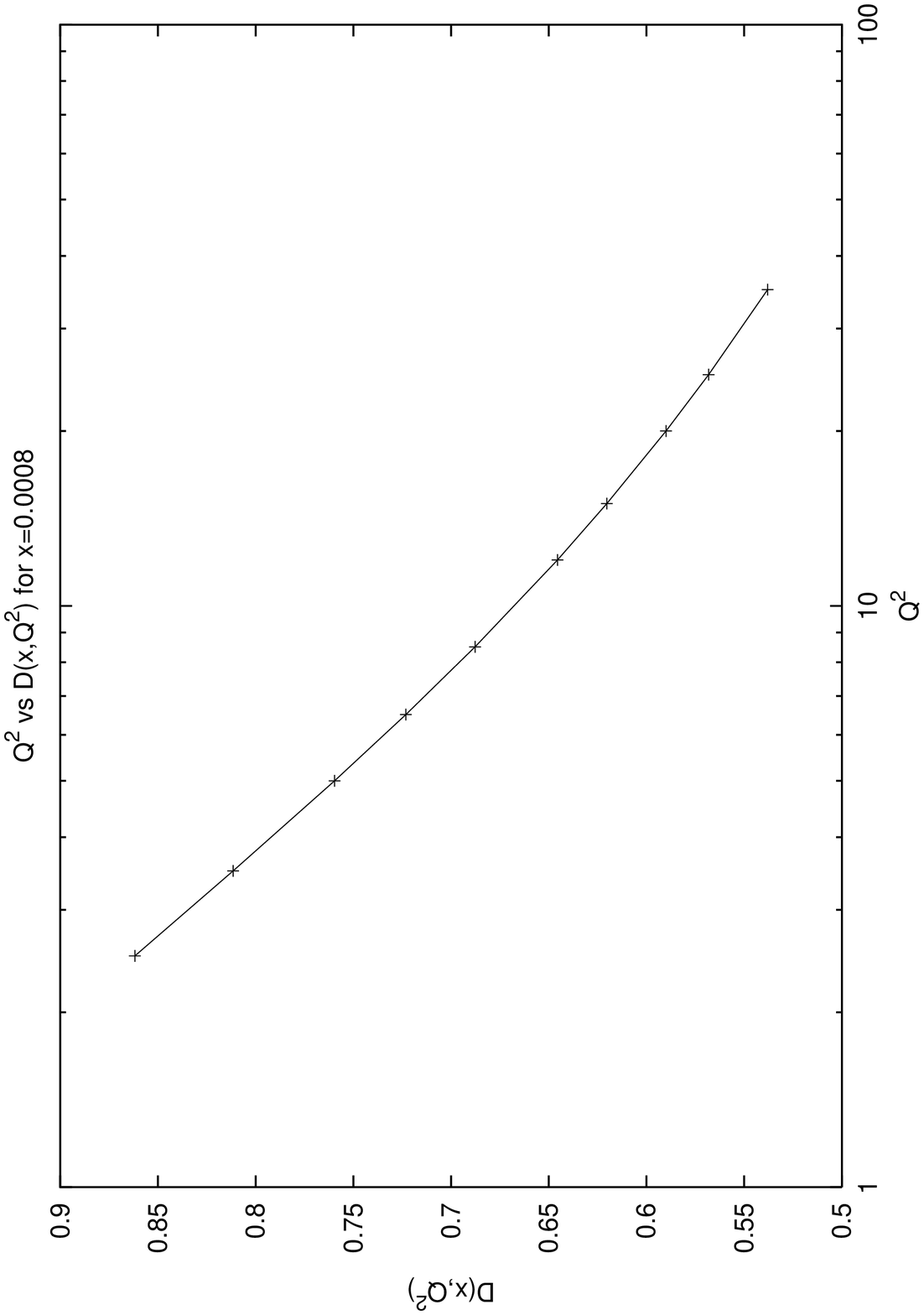}
\includegraphics[width=10cm]{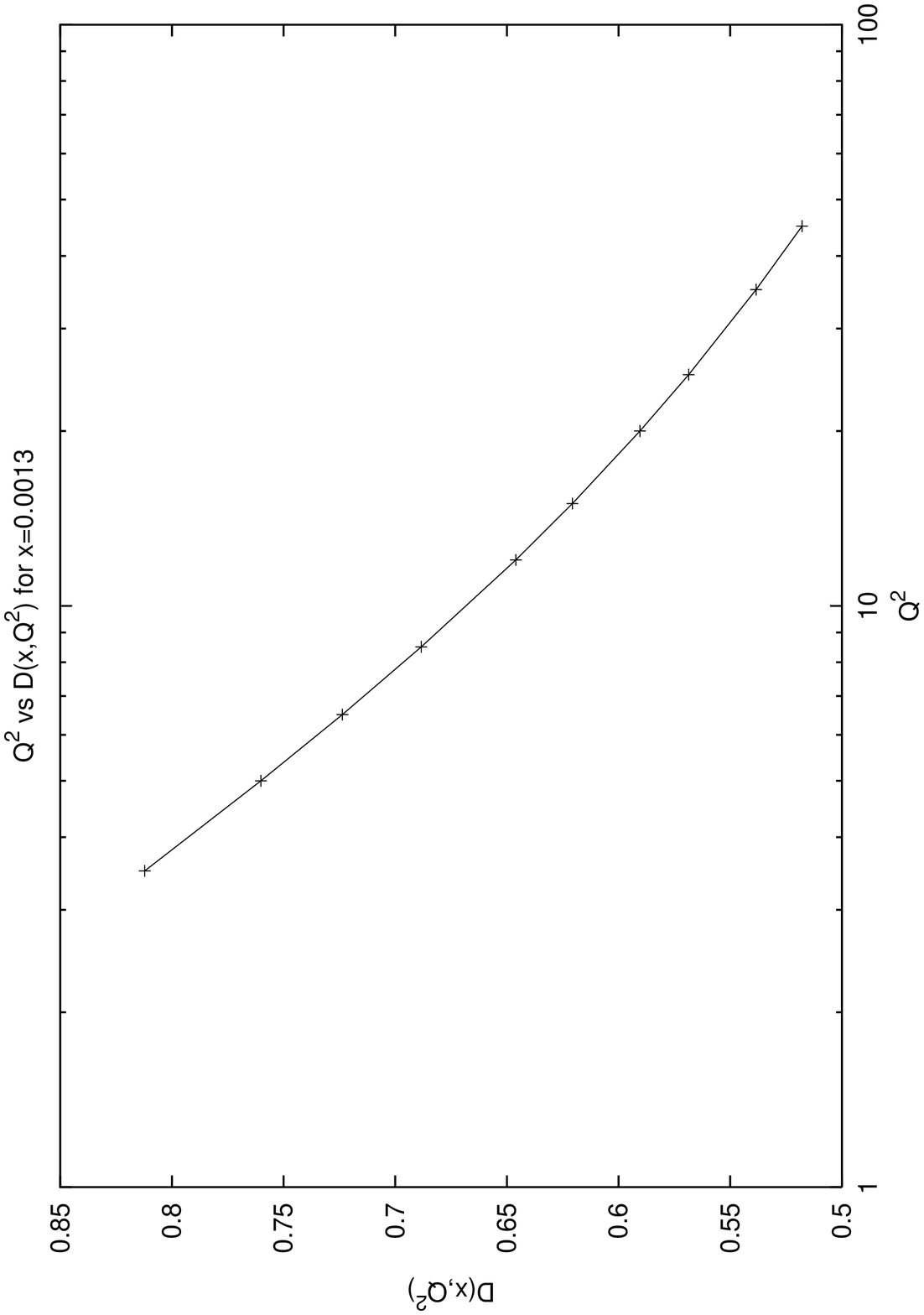}
}
\caption{$Q^{2}$ versus $D(x,Q^{2})$ for $x$=0.0008 and 0.0013}
\end{figure}
\begin{figure}[t]
\centerline{
\includegraphics[width=10cm]{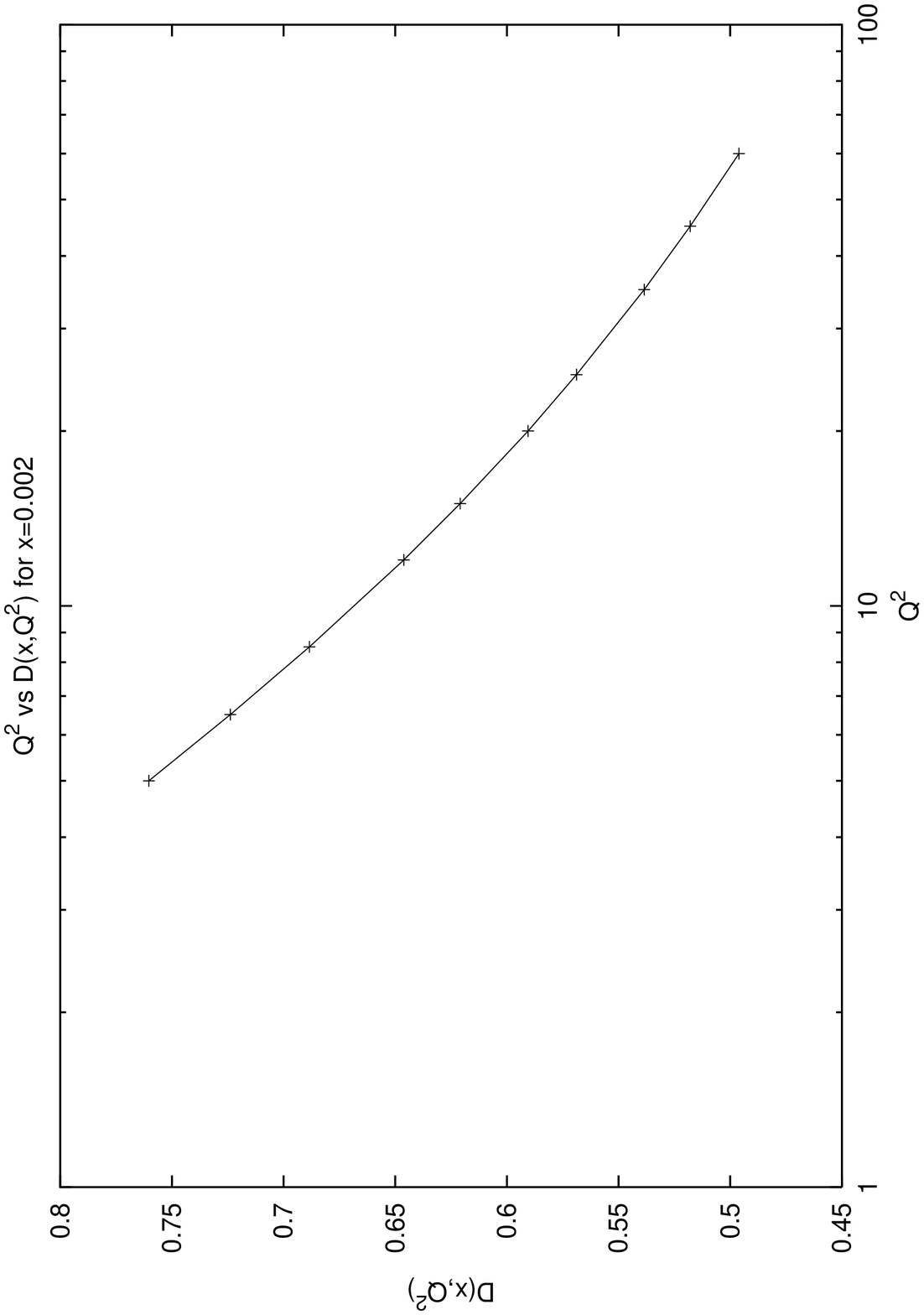}
\includegraphics[width=10cm]{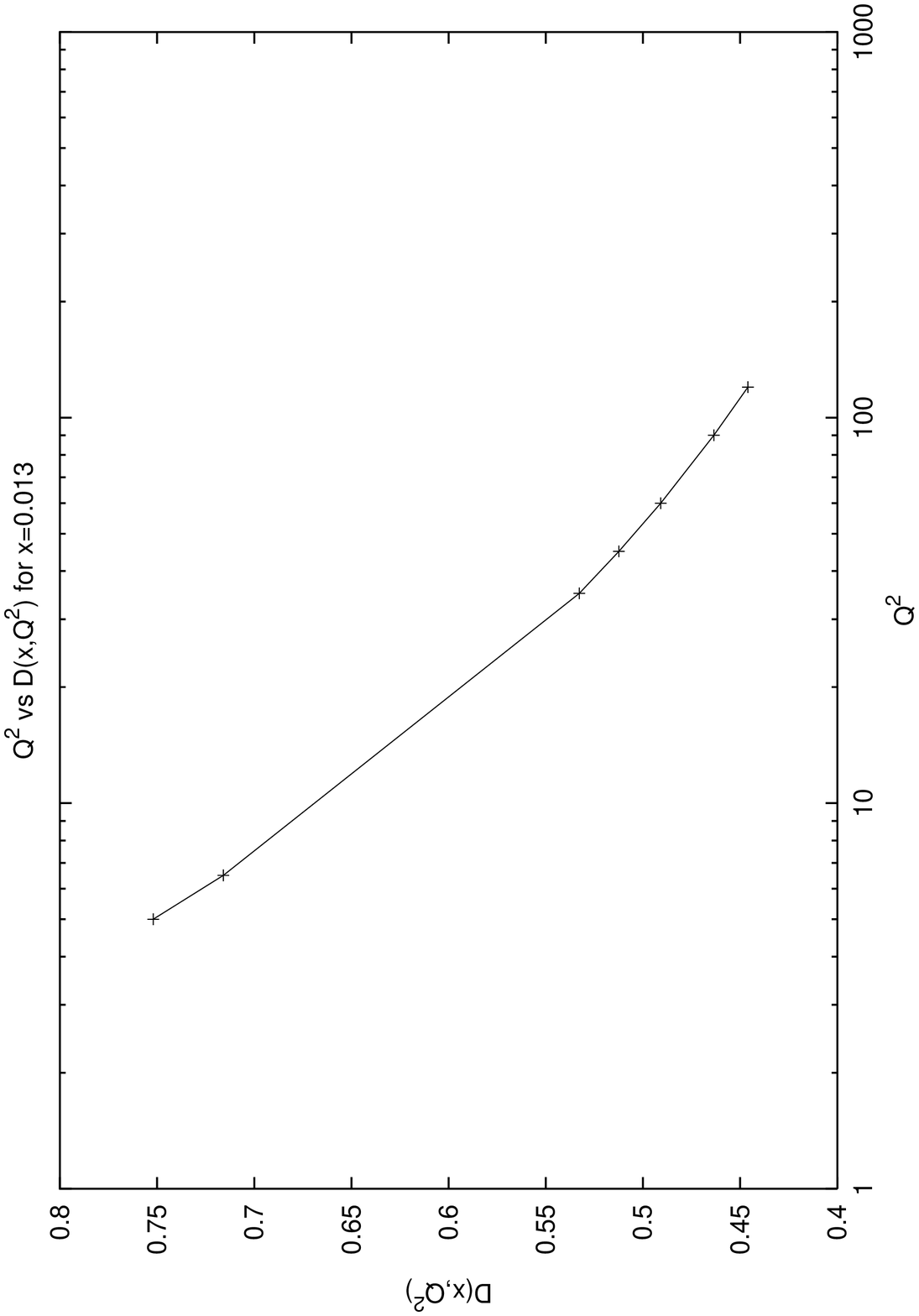}
}
\caption{$Q^{2}$ versus $D(x,Q^{2})$ for $x$=0.002 and 0.013}
\end{figure}
\end{document}